\documentclass[11pt,a4paper]{amsart}
\usepackage[utf8]{inputenc}

\usepackage{amsmath,amsthm,amssymb}

\usepackage[margin=2.5cm]{geometry}
\usepackage[nospace,noadjust]{cite}
\usepackage{relsize}
\usepackage{cite}
\usepackage{color}
\usepackage[dvipsnames]{xcolor}
\usepackage{hyperref, enumitem}
\hypersetup{
  colorlinks   = true, 
  urlcolor     = blue, 
  linkcolor    = Purple, 
  citecolor   = red 
}
\usepackage{stmaryrd}
\usepackage{comment}
\usepackage{mathtools}

\newcommand{\F}{\mathbb{F}}
\newcommand{\mC}{\mathcal{C}}
\newcommand{\mX}{\mathcal{X}}

\newcommand{\rk}{\mathrm{rk}}
\newcommand{\rowspan}{\mathrm{rowspan}}
\newcommand{\colspan}{\mathrm{colspan}}
\newcommand{\sss}{\mathrm{ss}}

\newcommand{\GL}{\mathrm{GL}}
\newcommand{\Fq}{\F_{q}}

\newcommand{\Tr}{\mathrm{Tr}}
\usepackage{etoolbox}
\renewcommand{\emph}[1]{\textbf{#1}}

\theoremstyle{definition}
\newtheorem{theorem}{Theorem}[section]
\newtheorem{definition}[theorem]{{{Definition}}}
\newtheorem{example}[theorem]{{{Example}}}
\newtheorem{remark}[theorem]{{{Remark}}}
\newtheorem{corollary}[theorem]{{{Corollary}}}
\newtheorem{proposition}[theorem]{{{Proposition}}}
\newtheorem{lemma}[theorem]{{{Lemma}}}

\title{Rank-metric intersecting codes}

\author{Martino Borello}
\address{Universit\'e Paris 8, Laboratoire de G\'eom\'etrie, Analyse et Applications, LAGA, Universit\'e Sorbonne Paris Nord, CNRS, UMR 7539, France}
\email{martino.borello@univ-paris8.fr}

\author{Martin Scotti}
\address{Aalborg University, Department of Mathematical Sciences, Denmark}
\email{msco@math.aau.dk}

\author{Lune Vila}
\address{Inria and LIX, École polytechnique, Institut Polytechnique de Paris, France}
\email{lune.vila@inria.fr}

\begin{document}

\begin{abstract}
Intersecting codes are codes in which any two nonzero codewords share a common nonzero coordinate, and they have been extensively studied due to their applications in communication theory, cryptography, and combinatorics. Recently, geometric interpretations have provided new insights into intersecting codes, both in the Hamming metric and, more recently, in the rank metric for linear codes.

In this paper, we introduce and study matrix rank-metric intersecting codes. By defining the support of a matrix as its row space, we extend the notion of intersecting codes to the matrix rank-metric setting. Using the slice space associated with the generator tensor of a rank-metric code, we derive a geometric characterization that is equivalent to the intersecting property. This characterization generalizes the known results for linear rank-metric intersecting codes and highlights both analogies and differences with the linear case. In particular, it yields bounds on the parameters of intersecting codes, one of which is sharp and differs from the one in the linear case, together with first constructions. We then consider codes that are closed under multiplication, i.e. matrix algebras. In this setting, the intersecting property can be read off from the algebraic structure, namely the Jacobson radical and the socle. This provides a simple and flexible way to construct many intersecting codes, including codes that do not satisfy the classical sufficient condition on the minimum distance.
\end{abstract}

\maketitle


\section*{Introduction}

Intersecting codes, first introduced in \cite{katona1983minimal, miklos1984linear}, are codes for which every two nonzero codewords have a nonzero coordinate in common. They have numerous applications, for instance for communication over an AND channel, for digital fingerprinting \cite{staddon2001combinatorial,blackburn2003frameproof}, for oblivious transfer \cite{brassard1996oblivious}, and they are related to several combinatorial structures such as $(2, 1)$-separating systems \cite{randriambololona20132}. In the binary case, they coincide with minimal codes, which have also been intensively studied \cite{AB, cohen2013minimal, ABNgeo}, and have been used by Massey to develop a secret sharing scheme \cite{massey1993minimal}. The theory of intersecting codes was further developed in various works such as \cite{retter1989intersecting, cohen1994intersecting,cohnen2003intersecting, randriambololona20132}. More recently, connections with additive combinatorics and factorization theory have also been explored \cite{SP,borello2025geometry}. In \cite{borello2025geometry}, the authors provide a geometric interpretation of intersecting codes in the Hamming metric: intersecting codes correspond to sets of points in the projective space that are not contained in the union of two hyperplanes. This allows various geometric investigations of intersecting codes.

\medskip

In the last two decades, rank-metric codes have become a central subject in coding theory, motivated by applications in areas such as network coding and post-quantum cryptography, as well as by their strong links with finite geometry. The geometric interpretation of linear rank-metric codes through $q$-systems \cite{randrianarisoa2020geometric} has proved to be a powerful tool for studying extremal classes of codes, including MRD and minimal linear rank-metric codes, and related structures \cite{sheekey2019linear,zini2021scattered,rankminimal_ABNR}. Linear intersecting codes in the rank metric have been introduced in \cite{bartoli2025linear} with the authors providing a geometric interpretation of intersecting codes in terms of $q$-systems, together with bounds on the parameters, constructions, and relations with other combinatorial structures. In particular, in \cite{bartoli2025linear} it is shown that linear rank-metric intersecting codes coincide with $2$-rank-frameproof codes and are $(2,1)$-separating, which suggests possible applications to digital fingerprinting in the context of network coding. Finally, in \cite{geometryofrankmetric}, the first author \textit{et al.} investigate the geometry of matrix rank-metric codes.

\medskip

Building on these previous works, our aim is to propose a definition of matrix rank-metric intersecting codes. In analogy with \cite{bartoli2025linear}, we define the support of a matrix to be its row space (one could equally consider the column space, which yields a similar theory by transposition). Our study follows two complementary approaches.

\medskip

The first one is geometric. Examining the third slice space of a generator tensor of the code, we derive a property of this space, namely column-hyperplane-indecomposability, which is equivalent to the code being row-intersecting. This property specializes to the one developed in \cite{bartoli2025linear} in the $\F_{q^m}$-linear case. It allows us to derive bounds on the parameters of intersecting codes, pointing out both the similarities and the differences with \cite{bartoli2025linear}. In particular, a column-nondegenerate row-intersecting $[m\times n,k]_q$ code satisfies $n\le 2m-1$, and this bound is sharp, whereas it can be improved to $2m-3$ in the $\F_{q^m}$-linear case. The geometric approach also provides first constructions, obtained by extending MRD codes.

\medskip

The second approach is algebraic. When the code is closed under multiplication, that is, when it is a matrix algebra, the row-intersecting property can be read off from its structure as a module over itself. We show that a row-intersecting algebra has a field as quotient by its Jacobson radical, and that the property is entirely determined by the socle of the algebra. In particular, every algebra with a simple socle is row-intersecting. This makes it easy to produce many intersecting codes, including codes that do not satisfy the natural sufficient condition $2d>n$ on the minimum distance, and to test the property efficiently. We conclude with a complete description of the intersecting codes $\F_q[B]$ generated by a single matrix.

\bigskip

\noindent \textbf{Outline:} Our work is organized as follows. In Section~\ref{Sec:prelim}, we give a brief introduction to rank-metric codes, tensors and the basic notions of representation theory that we need, and we introduce intersecting codes in this context, together with a first sufficient condition based on the minimum distance. In Section~\ref{Sec:results}, using the geometric interpretation of matrix rank-metric codes developed in \cite{geometryofrankmetric}, we characterize row-intersecting codes in terms of their slice spaces and derive bounds and constructions. In Section~\ref{sec:algebras}, we focus on codes that are matrix algebras and use their algebraic structure to characterize and construct row-intersecting codes.

\medskip

\section{Background and first properties} \label{Sec:prelim}

\subsection{Rank metric codes} We consider tensor products of the form $\F_q^{N_1}\otimes \F_q^{N_2}\otimes \F_q^{N_3}$,
whose elements are called $3$-tensors. The elements of this space can be represented as $3$-dimensional arrays. We can identify $\F_q^{N_1}\otimes \F_q^{N_2}\otimes \F_q^{N_3}$ with $\F_q^{N_1\times N_2\times N_3}$.
We introduce the following maps:
\begin{align*}
    m_1: &\ \F_q^{N_1\times N_2\times N_3} \times \F_q^{N_1\times s} \to \F_q^{s\times N_2\times N_3} \\ &(T,A_1) \mapsto m_1(T, A_1) := \sum_i (u_iA_1) \otimes v_i \otimes w_i,\\
    m_2: &\ \F_q^{N_1\times N_2\times N_3}\times \F_q^{N_2\times s} \to \F_q^{N_1\times s\times N_3} \\ &(T,A_2) \mapsto m_2(T, A_2) := \sum_i u_i \otimes (v_iA_2) \otimes w_i,\\
    m_3: &\ \F_q^{N_1\times N_2\times N_3} \times \F_q^{N_3\times s} \to \F_q^{N_1\times N_2\times s}\\& (T,A_3) \mapsto m_3(T, A_3) := \sum_i u_i \otimes v_i \otimes (w_iA_3),
\end{align*}
for any $T =\displaystyle\sum_i u_i \otimes v_i \otimes w_i \in\F_q^{N_1\times N_2 \times N_3}$.

\begin{definition}
    Let $T\in\F_q^{N_1 \times N_2\times N_3}$. For each $i\in\{1, 2, 3\}$, we define the \emph{$i$-th slice space} of $T$ to be the $\F_q$-span of $\{m_i(T,e_j) : 1 \leq j \leq N_i\}$, that is,
$$\sss_i(T) := \langle m_i(T,e_1),\ldots, m_i(T,e_{N_i})\rangle_{\F_q}.$$
\end{definition}

\begin{definition} 
An $[m\times n,k,d]_q$ \emph{rank-metric code} $\mC$ is a subspace of $\F_q^{m\times n}$ of \textbf{dimension} $k$  and \textbf{minimum distance} 
\[d:=\min\{\rk(A-B):A,B\in \mC, A\neq B\}=\min\{\rk(A):A\in \mC,A\neq 0\}.\]
A \emph{generator tensor} for a code $\mC\subseteq \F_q^{m\times n}$ is an element  $T\in\F_q^{k\times m\times n}$ such that $\mC=\sss_1(T)$. The \emph{dual} of a rank-metric code $\mC\subseteq \F_q^{m\times n}$ is the rank-metric code
$$\mC^\perp:=\{B:{\rm Tr}(BA^\top)=0,\ \forall A\in \mC\}.$$
\end{definition}

\begin{remark}
An $\F_q$-\emph{linear isometry} (with respect to the rank) of $\F_q^{m\times n}$ is a map $\varphi_{M,N}:\F_q^{m\times n}\to\F_q^{m\times n}$ sending $A$ to $MAN$, where $M\in \GL_m(\F_q)$ and $N\in \GL_n(\F_q)$. If $n=m$, transposition is also an isometry, but we avoid using that term in order to prevent confusion later. 
\end{remark}

\begin{definition}
Two $[m\times n,k,d]_q$ rank-metric codes $\mC,\mC'$ are \emph{equivalent} if there is an $\F_q$-linear isometry sending $\mC$ to $\mC'$. 
\end{definition}

\begin{definition}
The row-\emph{support} (respectively the column-\emph{support}) of a matrix $A\in \F_q^{m\times n}$ is $\sigma_r(A):=\rowspan(A)$ (respectively $\sigma_c(A):=\colspan(A)$).
\end{definition}

\begin{theorem}[Singleton-like bound, \cite{delsarte1978bilinear}]\label{thm:singleton}
Let $\mC$ be an $[m\times n,k,d]_q$ code. Then
\[
k\leq \max(m,n)\,\big(\min(m,n)-d+1\big).
\]
\end{theorem}

\begin{definition}
A code attaining the bound of Theorem~\ref{thm:singleton} is called a \emph{maximum rank distance} (\emph{MRD}) code.
\end{definition}

MRD codes exist for all parameters $q$, $m$, $n$ and $1\leq d\leq\min(m,n)$: the classical construction is due to Delsarte~\cite{delsarte1978bilinear} and, independently, Gabidulin~\cite{gabidulin1985theory} (see also~\cite{roth1991maximum}). We recall it in the vector setting. Let us assume $n\leq m$ (otherwise, one can transpose) and fix an ordered $\F_q$-basis $\Gamma=(\gamma_1,\ldots,\gamma_m)$ of $\F_{q^m}$ over $\F_q$. For $v=(v_1,\ldots,v_n)\in\F_{q^m}^n$, let $\Gamma(v)\in\F_q^{m\times n}$ be the matrix defined by
\[
v_j=\sum_{i=1}^m \Gamma(v)_{i,j}\,\gamma_i,\qquad 1\leq j\leq n,
\]
that is, the $j$-th column of $\Gamma(v)$ is the expansion of $v_j$ in the basis $\Gamma$. The map $\Gamma$ is $\F_q$-linear and injective, and $\rk(\Gamma(v))=\dim_{\F_q}\langle v_1,\ldots,v_n\rangle_{\F_q}$. Moreover, $\sigma_r(\Gamma(\lambda v))=\sigma_r(\Gamma(v))$ for every $\lambda\in\F_{q^m}^*$, and this row space does not depend on the choice of $\Gamma$.

\begin{definition}\label{def:gabidulin}
Let $1\leq s\leq n\leq m$ and let $g_1,\ldots,g_n\in\F_{q^m}$ be linearly independent over $\F_q$. The \emph{Gabidulin code} of dimension $s$ is
\[
\mathcal{G}_s(g_1,\ldots,g_n):=\left\{
\Gamma\big(f(g_1),\ldots,f(g_n)\big)\;:\;
f=\sum_{i=0}^{s-1}f_i\,x^{q^i},\ f_i\in\F_{q^m}
\right\}\subseteq\F_q^{m\times n}.
\]
\end{definition}

\begin{theorem}[\cite{delsarte1978bilinear,gabidulin1985theory}]\label{thm:gabidulin}
The code $\mathcal{G}_s(g_1,\ldots,g_n)$ is an MRD code with parameters $[m\times n,\,ms,\,n-s+1]_q$.
\end{theorem}

\begin{remark}
The proof of Theorem~\ref{thm:gabidulin} relies on the fact that a nonzero $q$-polynomial of $q$-degree at most $s-1$ has a kernel of $\F_q$-dimension at most $s-1$. Gabidulin codes are $\F_{q^m}$-linear. Other families of MRD codes, not necessarily equivalent to Gabidulin codes, are known, for instance the (generalized) twisted Gabidulin codes~\cite{lunardon2018generalized}.
\end{remark}

\subsection{Basic notions in representation theory} 

We will now recall some standard definitions and results that will be useful in Section \ref{sec:algebras} (we refer the reader to \cite{DrozdKirichenko} for their proofs and for further notions in representation theory). 

In all that follows, $\mathcal{A}$ will denote a finite-dimensional unital algebra over a field $\mathbb{K}$ and $\mathcal{M}$ a left module over this algebra. 

\begin{definition}
 A module $\mathcal{M}\neq \{0\}$ is said to be \textbf{simple} if its only submodules are $\{0\}$ and $\mathcal{M}$. A module is said to be \textbf{semisimple} if it is a direct sum of simple submodules.
\end{definition}

\begin{definition}
    The \textbf{Jacobson radical} of a module, denoted by ${\rm Rad}(\mathcal{M})$, is the intersection of all maximal submodules of $\mathcal{M}$. For the algebra $\mathcal{A}$, the radical ${\rm Rad}(\mathcal{A})$ is the intersection of all maximal (two-sided) ideals of $\mathcal{A}$.
\end{definition}

\begin{remark}\label{r1}
  For every module $\mathcal{M}$, the quotient $\mathcal{M}/{\rm Rad}(\mathcal{M})$ is semisimple. 
\end{remark}

\begin{definition}
    The \textbf{socle} ${\rm Soc}(\mathcal{M})$ of a module $\mathcal{M}$ is the sum of all its simple submodules.
\end{definition}

\begin{remark}
    It is well-known that
    \[{\rm Soc}(\mathcal{M})=\{m\in \mathcal{M}\ : \: {\rm Rad}(\mathcal{A})\cdot m=0\}\]
    and that the socle ${\rm Soc}(\mathcal{M})$    
    is the largest (with respect to inclusion) semisimple submodule of $\mathcal{M}$.
\end{remark}

\begin{definition}
Let $n=\dim_{\mathbb{K}}\mathcal{A}$ and fix a basis of $\mathcal{A}$, so
  that its elements are identified with row vectors of $\mathbb{K}^n$.
For
  $A\in\mathcal{A}$, let $M_A\in\mathbb{K}^{n\times n}$ denote the matrix of
  the right multiplication $X\mapsto XA$. The map    
    \[
\begin{aligned}
\rho \colon \mathcal{A} &\to \mathbb{K}^{n \times n} \\
A &\mapsto M_A
\end{aligned}
\]
is an injective morphism of $\mathbb{K}$-algebras; its image is called the
\textbf{right regular representation} of $\mathcal{A}$.
\end{definition}

\subsection{Rank intersecting codes}

We define now the main object of this paper.

\begin{definition}
A code $\mC\subseteq \F_q^{m\times n}$ is row-\emph{intersecting} (respectively column-\emph{intersecting}) if every pair of nonzero \emph{codewords} $A,B\in \mC$ has non-trivially intersecting row-supports (respectively column-supports), that is for all
$A,B\in \mC\setminus\{0\}$,    \[\sigma_r(A)\cap \sigma_r(B)\neq \{0\}\] (respectively for all $A,B\in \mC\setminus\{0\}, \   \sigma_c(A)\cap \sigma_c(B)\neq \{0\}$).
\end{definition}

\begin{remark}
Every $\F_q$-linear rank-metric code of dimension $1$ is trivially both row- and column-intersecting, since any two nonzero codewords are $\F_q$-proportional. Note however that this is no longer true for codes which are linear over an extension field: for instance, an $\F_{q^m}$-linear code of dimension $1$ is always row-intersecting, but not necessarily column-intersecting.
\end{remark}

\begin{remark} \label{rem1}
Being intersecting is invariant under equivalence. Moreover, the transposition sends row-intersecting codes to column-intersecting ones.    
\end{remark}

\begin{example}
The $[4\times 3,8,2]_2$ code generated by the following matrices
\[\left[\begin{smallmatrix}
1&0&0\\
0&1&0\\
0&0&1\\
0&0&0
\end{smallmatrix}\right],
\left[\begin{smallmatrix}
0&0&0\\
1&0&0\\
0&1&0\\
0&0&1
\end{smallmatrix}\right],
\left[\begin{smallmatrix}
0&0&1\\
0&0&1\\
1&0&0\\
0&1&0
\end{smallmatrix}\right],
\left[\begin{smallmatrix}
0&1&0\\
0&1&1\\
0&0&1\\
1&0&0
\end{smallmatrix}\right],
\left[\begin{smallmatrix}
1&0&1\\
0&0&1\\
0&1&0\\
0&0&0
\end{smallmatrix}\right],
\left[\begin{smallmatrix}
0&0&0\\
1&0&1\\
0&0&1\\
0&1&0
\end{smallmatrix}\right],
\left[\begin{smallmatrix}
0&1&0\\
0&1&0\\
1&0&1\\
0&0&1
\end{smallmatrix}\right],
\left[\begin{smallmatrix}
0&0&1\\
0&1&1\\
0&1&0\\
1&0&1
\end{smallmatrix}\right]\]
is row-intersecting but not column-intersecting. The $[4\times 3,4,3]_2$ code genera\-ted by the first four matrices is both row-intersecting and column-intersecting. Both codes are Gabidulin codes (see Definition~\ref{def:gabidulin}).
\end{example}

The following result gives a sufficient condition which is easy to check and which allows one to easily construct examples.

\begin{proposition}\label{prop:min}
Let $\mC$ be an $[m\times n,k,d]_q$ code. If $2d>n$, then $\mC$ is row-intersecting. If $2d>m$, then $\mC$ is column-intersecting.   
\end{proposition}

\begin{proof}
    This follows immediately from Grassmann's formula, because
    \[\dim (\sigma_r(A)\cap \sigma_r(B))=\dim \sigma_r(A)+\dim \sigma_r(B)-\dim (\sigma_r(A)+\sigma_r(B))>\frac{n}{2}+\frac{n}{2}-n=0.\]
\end{proof}

\begin{remark}\label{rmk:constructions}
Note that Proposition \ref{prop:min} allows constructions of rank-metric intersecting codes: in fact, it is enough to consider known constructions of MRD codes (Gabidulin codes, see Theorem~\ref{thm:gabidulin}, or their generalizations) with parameters satisfying the hypothesis of the proposition. Let us also remark that this gives examples of rank-metric intersecting codes which are not necessarily linear. However, we stress that the sufficient condition of Proposition \ref{prop:min} only involves the minimum rank distance of $\mC$, i.e.\ it is a property of its rank weight distribution, whereas being row- or column-intersecting is a structural property, depending on how the row spaces (resp.\ column spaces) of the codewords sit inside $\F_q^n$ (resp.\ $\F_q^m$). In particular, this condition is not necessary, and codes with $2d\le n$ (resp.\ $2d\le m$) require different arguments.
\end{remark}

Roughly speaking, a rank-metric code is row-intersecting if and only if it is intersecting in the Hamming sense with respect to every choice of coordinates in $\F_q^n$, and similarly for column-intersecting codes and $\F_q^m$. In order to make this statement precise, we introduce the following notion.

\begin{definition}
A rank-metric code $\mC\subseteq \F_q^{m\times n}$ is said to be \emph{Hamming row-intersecting} (respectively \emph{Hamming column-intersecting}) if for any $A, B \in \mC \setminus \{0\}$, there exists a coordinate $i$ for which the $i$-th rows (respectively the $i$-th columns) of $A$ and $B$ are both nonzero.
\end{definition}

\begin{theorem}
A rank-metric code $\mC$ is row-intersecting (respectively column-intersecting) if and only if, for all $N\in \GL_n(\F_q)$ (respectively $M\in \GL_m(\F_q)$), $\varphi_{I_m,N}(\mC)$ (respectively $\varphi_{M,I_n}(\mC)$) is Hamming column-intersecting (respectively Hamming row-intersecting).
\end{theorem}

\begin{proof}
Assume that for some $N \in \GL_n(\F_q)$, the code $\varphi_{I_m,N}(\mC)$ is not Hamming column-intersecting. Then there exist $2$ codewords $A,B$ such that $AN$ and $BN$ do not share a nonzero column. Hence $\sigma_r(AN)\cap \sigma_r(BN)=\{0\}$, which implies $\sigma_r(A)\cap \sigma_r(B)=\{0\}$. Therefore the code $\mC$ is not row-intersecting.

Conversely, assume that there exist two codewords $A,B \in \mC$ such that $\sigma_r(A) \cap \sigma_r(B) = \{0\}$. Then there exists $N \in \GL_n(\F_q)$ such that $AN$ and $BN$ have no coordinate $i$ for which both $AN$ and $BN$ have a nonzero column. Hence $\varphi_{I_m,N}(\mC)$ is not Hamming column-intersecting.
\end{proof}

\medskip

\section{A geometric approach: the third slice space of a row-intersecting code} \label{Sec:results}

We will now concentrate on row-intersecting codes. We do this for the sake of brevity, and an entirely analogous argument applies to column-intersecting codes as well.

For a code $\mC\subseteq \F_q^{m\times n}$ of dimension $k$, generated by $T$, let us consider
\[\mX = \sss_3(T)\subseteq \F_q^{k\times m}.\]
\noindent \textbf{From now on, we will assume, without loss of generality, that $\dim \mX=n$} (which means that the code $\mC$ is column-nondegenerate): if this is not the case, $\mC$ can be isometrically embedded in a space of matrices $\F_q^{m\times n'}$, with $n'\leq n$, such that $\dim \mX=n'$.

\begin{definition}
Let $l\in \{1,\ldots,k-1\}$, $r\in \{1,\ldots,m-1\}$, $L\in \F_q^{l\times k}$ of full rank, $R\in \F_q^{m\times r}$ of full rank. The \emph{left puncturing} of $\mX$ with respect to $L$ is 
\[_L\mX=\{LM:M\in \mX\}\subseteq \F_q^{l\times m}.\]
The \emph{right puncturing} of $\mX$ with respect to $R$ is 
\[\mX_R=\{MR:M\in \mX\}\subseteq \F_q^{k\times r}.\]
Moreover, for a subspace $S\subseteq \F_q^k$, we define
$$\mX_c(S) := \{M \in \mX : \colspan(M) \subseteq S \}$$
and for a subspace $S\subseteq \F_q^m$, we define
$$\mX_r(S) := \{M \in \mX : \rowspan(M) \subseteq S \}$$
The \emph{left-shortening} of $\mX$ with respect to $L$ is $_L\mX_c(\rowspan(L))\subseteq \F_q^{l\times m}$. The \emph{right-shortening} of $\mX$ with respect to $R$ is $\mX_r(\colspan(R))_R\subseteq \F_q^{k\times r}$.
\end{definition}

\begin{definition}
We say that a (slice) space is \emph{column-hyperplane-decom\-po\-sa\-ble} if there exist two hyperplanes $H_{1},H_{2}$ of $\F_q^k$ such that
$$\mX = \mX_c(H_{1}) + \mX_c(H_{2}).$$
We say that a slice space is \emph{column-hyperplane-indecomposable} if it is not column-hyperplane-decomposable.
\end{definition}

\begin{theorem}
Let $\mC\subseteq \F_q^{m\times n}$ be a rank-metric code of dimension $k$ with generator tensor $T$ and let $\mX=\sss_3(T)\subseteq \F_q^{k\times m}$ be the third slice space of $T$. Then 
\[\mC\text{ is row-intersecting} \ \ \Longleftrightarrow \ \ \mX \text{ is column-hyperplane-indecomposable.}\]
\end{theorem}

\begin{proof}
As shown in \cite[Lemma~3.5]{geometryofrankmetric}, for all $u\in \F_q^k$,
\begin{equation}\label{eq:support}
\sigma_r(m_1(T,u))^\perp=\psi^{-1}(\mX_c(\langle u\rangle_{\F_q}^\perp))    
\end{equation}
where
$$\psi:\F_q^n\to \mX, \ \ \  v\mapsto m_3(T,v).$$
Here the duality is with respect to the standard inner product in the corresponding spaces. The slice space
$\mX$ is column hyperplane-decomposable if and only if there exists $u,u'\in \F_q^k\setminus\{0\}$ such that 
$$\mX=\mX_c(\langle u\rangle_{\F_q}^\perp)+\mX_c(\langle u'\rangle_{\F_q}^\perp).$$
This is equivalent to 
$$\F_q^n=\psi^{-1}(\mX_c(\langle u\rangle_{\F_q}^\perp))+\psi^{-1}(\mX_c(\langle u'\rangle_{\F_q}^\perp))=\sigma_r(m_1(T,u))^\perp+\sigma_r(m_1(T,u'))^\perp,$$
which happens if and only if
$$\{0\}=(\sigma_r(m_1(T,u))^\perp+\sigma_r(m_1(T,u'))^\perp)^\perp=\sigma_r(m_1(T,u))\cap \sigma_r(m_1(T,u')).$$
The last condition means that $\mC$ is not row-intersecting.
\end{proof}

\begin{remark}\label{rmk:Fqmlinear}
Let $\mC=\{\Gamma(v):v\in C\}$, where $C\subseteq\F_{q^m}^n$ is an $\F_{q^m}$-linear code of dimension $k'$, so that $k=k'm$. For every $v\in C$, the row space $\sigma_r(\Gamma(v))\subseteq\F_q^n$ is the rank support of $v$ in the sense of \cite{bartoli2025linear}; hence $\mC$ is row-intersecting if and only if $C$ is intersecting in the sense of \cite{bartoli2025linear}. In that paper, it is shown that this holds if and only if the associated $q$-system $U\subseteq\F_{q^m}^{k'}$ is not $2$-spannable, that is, there are no two $\F_{q^m}$-hyperplanes $H_1,H_2$ of $\F_{q^m}^{k'}$ such that $U=(U\cap H_1)+(U\cap H_2)$ \cite[Theorem~3.4]{bartoli2025linear}. Therefore, in the $\F_{q^m}$-linear case, the column-hyperplane-indecomposability of the third slice space $\mX$ is equivalent to the fact that $U$ is not $2$-spannable, and the theorem above recovers the geometric characterization of \cite{bartoli2025linear}. Note, however, that column-hyperplane-decomposability involves all pairs of $\F_q$-hyperplanes of $\F_q^{k}$, while $2$-spannability only involves $\F_{q^m}$-hyperplanes of $\F_{q^m}^{k'}$; the equivalence of the two conditions reflects the fact that, for $\F_{q^m}$-linear codes, the row space of a codeword is invariant under multiplication by $\F_{q^m}^*$.
\end{remark}

\begin{proposition}\label{prop:sufficient}
If, for every hyperplane $H \subseteq \F_q^k$, 
$$\dim\mX_c(H) < \frac{n}{2},$$
then $\mX$ is column-hyperplane-indecomposable.
\end{proposition}

\begin{proof}
For any two hyperplanes $H_1,H_2\subseteq\F_q^k$,
\[
\dim(\mX_c(H_1)+\mX_c(H_2))\leq \dim\mX_c(H_1)+\dim\mX_c(H_2)<\dim\mX,
\]
so $\mX\neq\mX_c(H_1)+\mX_c(H_2)$.
\end{proof}

\begin{theorem}\label{thm:lowerbound}
If $\dim \mX\leq 2\lfloor (k-1)/m\rfloor$, then $\mX$ is column-hyperplane-decomposable.
\end{theorem}

\begin{proof}
Let $X_1, \dots, X_n$ be an $\Fq$-linear basis of the slice space $\mX$. For each $1 \leq i \leq n$ write $V_{i} := \colspan(X_i)^{\perp}$, which is a vector subspace of $\Fq^k$ of codimension at most $m$.
Setting $j = \lfloor\frac{k-1}{m}\rfloor$, we have $W_1 = V_1 \cap \dots \cap V_j \neq \{0\}$. Indeed, the codimension of $W_1$ is at most $mj \leq k-1$ by Grassmann's formula. Therefore there exists a nonzero vector $w_1 \in W_1 \setminus \{ 0 \}$. Similarly, we set $W_2 = V_{j+1} \cap \dots \cap V_{n}$. Since $n = \dim \mX \leq 2j$, we also have $W_{2} \neq \{0\}$, so that we may also choose a nonzero vector $w_2 \in W_2 \setminus \{ 0 \}$.

Defining the hyperplanes $H_1 = \langle w_1 \rangle_{\F_q}^{\perp}$ and $H_2 = \langle w_2 \rangle_{\F_q}^{\perp}$, for all $1 \leq i \leq j$, we have $X_i \in \mX_c(H_1)$, while for all $j+1 \leq i \leq n$, we have $X_i \in \mX_c(H_2)$. Therefore
$$\mX = \mX_c(H_1) + \mX_c(H_2),$$
meaning that $\mX$ is column-hyperplane-decomposable, as claimed.
\end{proof}

\begin{lemma}
For every subspace $S\subseteq \F_q^k$ of codimension $s$,
$$\dim \mX_c(S)\geq n -sm.$$
\end{lemma}

\begin{proof}
For every subspace $S\subseteq \F_q^k$ of codimension $s$ we have
$$\mX_c(S)=\mX\cap(S\otimes \F_q^m)$$
so that
\begin{align*}
\dim \mX_c(S)&=\dim (\mX\cap(S\otimes \F_q^m))\\&=\dim \mX+\dim (S\otimes \F_q^m)-\dim (\mX+(S\otimes \F_q^m))\\
& \geq \dim \mX+(k-s)m-km\\
& =\dim \mX -sm \\
& = n - sm.
\end{align*} 
\end{proof}

\begin{theorem}\label{thm:md}
    If $\mX$ is column-hyperplane-indecomposable, then $$\dim \mX_c(H)\leq n-\frac{k}{m}$$ for every hyperplane $H\subseteq \F_q^k$. Hence, if $\mC$ is a $[m\times n,k,d]_q$ row-intersecting code with $\dim \mX=n$, then $m\cdot d\geq k$.
\end{theorem}

\begin{proof}
    Let $\dim \mX_c(H)=r$ and $X_1,\ldots, X_r$ a basis of $\mX_c(H)$. Let $X_{r+1},\ldots,$ $X_{n}$ be $n-r$ linearly independent matrices in $\F_q^{k\times m}$ that, with the matrices $X_1,\ldots,X_r$, form a basis for $\mX$. If $m\cdot (n-r)\leq k-1$, there exists a hyperplane $H'$ such that 
    $X_{r+1},\ldots,X_{n}$ belong to $\mX_c(H')$, so that $\mX$ is column-hyperplane-decomposable. For the second statement, let $u\in\F_q^k\setminus\{0\}$. By \eqref{eq:support}, since $\psi$ is an isomorphism, $\rk(m_1(T,u))=\dim\sigma_r(m_1(T,u))=n-\dim\mX_c(\langle u\rangle_{\F_q}^\perp)\geq k/m$. Since every nonzero codeword is of the form $m_1(T,u)$, we get $d\geq k/m$.
\end{proof}

\begin{theorem}\label{thm:upperbound}
$$\mX \text{ column-hyperplane-indecomposable} \Rightarrow \dim\mX \leq 2m - 1.$$
\end{theorem}

\begin{proof}
Let $H_1,H_2$ be two hyperplanes.  
\begin{align*}
\dim (\mX_c(H_1)+\mX_c(H_2))&=\dim \mX_c(H_1)+\dim \mX_c(H_2) -\dim \mX_c(H_1\cap H_2)\\
&\geq 2(\dim \mX-m)-\dim \mX_c(H_1\cap H_2)
\end{align*}
So that, if 
$\dim \mX_c(H_1\cap H_2)\leq\dim\mX-2m$,
then $\mX$ is column-hyperplane-decomposable. 

Suppose now $\dim \mX=2m$. Via an isomorphism $\F_q^m\simeq \F_{q^m}$, we may consider $\mX$ to be an $\F_q$-subspace of $\F_{q^m}^k$, which corresponds to a set of points in ${\rm PG}(k-1,q^m)$ of cardinality at most $\frac{q^{2m}-1}{q-1}$. Such a set is too small to meet all $\F_{q^m}$-subspaces of ${\rm PG}(k-1,q^m)$ of codimension $2$ (see  \cite[Theorem 2]{bose1966characterization}). Now, such subspaces correspond, via the isomorphism mentioned before, to the subspaces of $\F_q^{k\times m}$ of type $S\otimes \F_q^m$, with $S$ a subspace of codimension $2$. This means that there exists $S$ of codimension $2$ such that $\mX_c(S)=\{0\}$. Let $H_1,H_2$ such that $S=H_1\cap H_2$. Then $\dim \mX_c(H_1\cap H_2)=0=2m-2m$, so that $\mX$ is column-hyperplane-decomposable.

If $\dim\mX>2m$, it is enough to pick a subspace of $\mX$ of dimension $2m$ and repeat the argument, as in \cite{bartoli2025linear}.
\end{proof}

\begin{remark}
 Note that the argument used in \cite{bartoli2025linear} to get $2m-3$ instead of $2m-1$ does not work in this setting. Moreover, the bound is sharp: let $\mC$ be an MRD code of parameters $[m\times(2m-1),2m-1,m]_q$. It is row-intersecting by Proposition~\ref{prop:min}, since $2d=2m>2m-1$, and the same holds for any of its subcodes. Its third slice space has dimension $2m-1$: otherwise, $\mC$ would be isometric to a code of $\F_q^{m\times(2m-2)}$ with the same dimension and minimum distance, contradicting Theorem~\ref{thm:singleton}. Finally, in the $\F_{q^m}$-linear case, the upper bound has been recently refined in \cite{borello2026existence}, by showing that the dual of the $q$-system associated with an intersecting code is an evasive subspace. It is not clear whether this machinery extends to the $\F_q$-linear setting by passing to the dual, since it heavily relies on the $\F_{q^m}$-linear structure, for instance through the count of the $2$-dimensional $\F_{q^m}$-subspaces through a given point.
\end{remark}

\begin{remark} 
Theorem \ref{thm:lowerbound} and \ref{thm:upperbound} summarize as follows, in coding-theore\-ti\-cal terms: let $\mC\subseteq \F_q^{m\times n}$ be a rank-metric code of dimension $k$ with generator tensor $T$ and let $\mX=\sss_3(T)\subseteq \F_q^{k\times m}$ be the third slice space of $T$. If $\mC$ is row-intersecting, then 
$$2\lfloor (k-1)/m\rfloor< \dim \mX<2m.$$
\end{remark}

In the theorem, we generalize a result of \cite{bartoli2025linear} to matrix intersecting codes. The result shows that linearity over the extended field is actually superfluous for extending MRD codes and obtaining (row)-intersecting codes. The result is formulated in terms of the third slice space.

\begin{theorem}
Let $1\leq h\leq \left\lfloor \frac{m+1}{2}\right\rfloor$. There exists a column-hyperplane-indecomposable $\mX\subseteq \F_q^{hm\times m}$ with 
$$ \dim \mX\in \{m,\ldots,2m - 2h+1\}.$$
\end{theorem}
\begin{proof}
Let $\mX'\subseteq \F_q^{hm\times m}$ such that $\dim \mX'_c(H)\leq h-1$ for every hyperplane $H\subseteq\F_q^{hm}$ (which corresponds to a square MRD code -- e.g. a Gabidulin code -- of parameters $[m\times m,hm,m-h+1]_q$, see \cite{geometryofrankmetric}). Let $\mX''$ be an $\Fq$-linear subspace of $\F_q^{hm\times m}$ of dimension $r\leq (hm-1)m$ and such that $\mX'\cap \mX''=\{0\}$. 
Set $\mX = \mX' \oplus \mX''$. Then $\dim \mX = m + r$.
We want to prove that $\mX$ is column-hyperplane-indecomposable whenever $r\leq m-2h+1$.
Suppose $r\leq m-2h+1$ and assume that there exists a hyperplane $H$ such that $\dim \mX_c(H) \geq \frac{1}{2}\cdot \dim \mX$. Then
\begin{align*}
\dim \mX'_c(H) & = \dim(\mX'\cap( H\otimes\F_q^m)) =\dim(\mX\cap \mX'\cap( H\otimes\F_q^m))  \\& = \dim\mX_c(H)+ \dim\mX' - \dim(\mX_c(H) + \mX')\\ 
& \geq \dim\mX_c(H) + \dim\mX' - \dim\mX\\ & \geq \frac{m+r}{2} + m - (m + r) =\frac{m - r}{2}\geq \frac{m-(m-2h+1)}{2}=h-\frac{1}{2},
\end{align*}
from which we get $\dim \mX'_c(H)\geq h$, that contradicts the assumption above. It follows that $\dim\mX_c(H)<\frac{1}{2}\cdot \dim \mX$ for every hyperplane $H$, so that $\mX$ is column-hyperplane-indecomposable by Proposition~\ref{prop:sufficient}.
\end{proof}

\begin{remark}
The above result, together with Remark~\ref{rmk:constructions}, guarantees the existence of (row)-intersecting codes of length between $m$ and $2m-2h+1$ and dimension $hm$, for $1 \leq h \leq \left\lfloor \frac{m+1}{2} \right\rfloor$. 
Clearly, dimensions that are not multiples of $m$ can also be obtained by considering subcodes.
\end{remark}

To conclude this section, we want to translate the condition of being column-hyper\-plane-indecomposable in terms of the dual of the slice space. Let us reformulate in our terms the duality shortening-puncturing which is proved in many previous works on rank-metric codes (see for example \cite{byrne2017covering}).

\begin{theorem}
Let $\mX\subseteq \F_q^{k\times m}$ and $L\in \F_q^{l\times k}$ of full rank.
$$(_{L}\mX)^\perp= {_A(}\mX^\perp)_c(\rowspan(L)),$$
where $A \in \Fq^{l \times k}$ is such that $A L^{t} = I_l$.
\end{theorem}
\begin{proof}
Let us first examine $(_{L}\mX)^\perp$. We have
\begin{align*}
(_{L}\mX)^\perp &= \{ N \mid \forall M \in \mX, \quad \Tr(N^{t}LM) = 0 \} \\
&= \{ N \mid \forall M \in \mX, \quad \Tr(N^{t}(L^{t})^{t}M) = 0 \} \\
&= \{ N \mid \forall M \in \mX, \quad \Tr((L^{t}N)^{t}M) = 0 \}
\end{align*}
Now note that for any $N \in (_{L}\mX)^\perp \subseteq \Fq^{l \times m}$ we have $$\colspan (L^{t}N) \subseteq \colspan(L^{t}) = \rowspan(L).$$ Note also that $\forall P \in \Fq^{k \times m}$ such that $\colspan(P) \subseteq \rowspan(L)$, there exists $N \in \Fq^{l \times m}$ such that $P = L^{t}N$. Therefore we can write
$$\mX^\perp_c(\rowspan(L)) = \{ P \mid \forall M \in \mX, \ \Tr(P^{t}M) = 0 \text{ and } \colspan(P) \subseteq \rowspan(L) \},$$
and so
\begin{align*}
(_{L}\mX)^\perp &= \{ N \mid \forall M \in \mX, \quad \Tr((L^{t}N)^{t}M) = 0 \} \\
&= \{ AL^{t}N \mid \forall M \in \mX, \quad \Tr((L^{t}N)^{t}M) = 0 \} \\
&= \{ AP \mid \forall M \in \mX, \quad \Tr(P^{t}M) = 0 \text{ and } \colspan(P) \subseteq \rowspan(L) \}
\\&= {_A(}\mX^\perp)_c(\rowspan(L)),
\end{align*}
which completes the proof.
\end{proof}

\begin{corollary}
Let $\mX\subseteq \F_q^{k\times m}$. The following conditions are equivalent:
\begin{itemize}
    \item[{\rm a)}] $\mX$ is column-hyperplane-decomposable;
    \item[{\rm b)}] There exist two full-rank matrices $L_1,L_2\in \F_q^{(k-1)\times k}$ such that 
$$\mX=\mX_c(\rowspan(L_1))+\mX_c(\rowspan(L_2));$$
\item[{\rm c)}] There exist two full-rank matrices $L_1,L_2\in \F_q^{(k-1)\times k}$ such that
$$\mX^{\perp} = \mX_c(\rowspan(L_1))^{\perp} \cap \mX_c(\rowspan(L_2))^{\perp};$$
\item[{\rm d)}] There exist two full-rank matrices $L_1,L_2\in \F_q^{(k-1)\times k}$ such that $$_{A_1}\mX=(_{L_1}\mX^\perp)^\perp+_{A_1}\mX_c(\rowspan(L_2)),$$
where $A_1$ is such that $A_1 L_1^{t} = I_{k-1}$.
\end{itemize}
\end{corollary}

\begin{remark}
The characterizations above provide a dual description of column-hyperplane-indecomposable spaces. Understanding how to exploit them, either for constructions or for bounds, as was done via duality in the $\F_{q^m}$-linear case \cite{borello2026existence}, is an interesting direction for future research.
\end{remark}

\medskip

\section{An algebraic approach: rank intersecting modules} \label{sec:algebras}

In the previous section, we have considered the general case: linear spaces of matrices of any size and without any further algebraic structures. In the present section, we focus on the special case where $m=n$ (i.e. all the matrices are square) and we consider rank-metric codes $\mC\subseteq \mathbb{F}_q^{n\times n}$ closed under multiplication, i.e., matrix algebras. We shall see how this additional property allows us to construct intersecting codes in an efficient and quite simple manner. \\

Throughout this section, let $\mathcal{C}$ be a finite dimensional unital algebra, seen as a left module over itself.

\begin{remark}
    As mentioned before, we will be focusing on the rowspan rather than on the columnspan. Let us note that Remark \ref{rem1} no longer holds here because the added structure does not allow us to transpose as easily. However, considering right-modules instead of left ones yields all the analogous results for column-intersecting codes.
\end{remark}

\begin{definition}
    An equality $$1_\mathcal{C}=e_1+\ldots+e_r,$$ where $e_1,\ldots,e_r\in\mathcal{C}$ are pairwise orthogonal idempotents (that is, $e_i^2=e_i$ and $e_i\neq 0$ for all $i$, and $e_ie_j=0$ for all $i\neq j$) is called a \textbf{decomposition of the identity}.
\end{definition}

We now show how these algebraic properties are related to the intersecting property of the code. 

\begin{lemma}\label{lemma:orth-idemp}
    Let $e,f\in\mathbb{F}_q^{n\times n}$ be idempotents such that $fe=0$.
    Then $\sigma_r(e)\cap\sigma_r(f)=\{0\}$.
\end{lemma}

\begin{proof}
    Since $e^2=e$, we have $\sigma_r(e)=\{v\in\mathbb{F}_q^n : ve=v\}$:
    indeed, if $v=xe$ then $ve=xe^2=v$, and the converse is clear.
    The same holds for $f$. If $v\in\sigma_r(e)\cap\sigma_r(f)$, then
    \[
    v=ve=(vf)e=v(fe)=0.
    \]
\end{proof}

\begin{lemma}\label{lemma:simple-rowspan}
    Let $\mathcal{S}\subseteq\mathcal{C}$ be a simple left $\mathcal{C}$-submodule. Then all nonzero elements of $\mathcal{S}$ have the same row space. In particular, $\mathcal{S}$ is row-intersecting.
\end{lemma}

\begin{proof}
    Let $A,A'\in\mathcal{S}\setminus\{0\}$. Since $A=1_{\mathcal{C}}A\in\mathcal{C}A\subseteq\mathcal{S}$, the submodule $\mathcal{C}A$ is nonzero, hence $\mathcal{C}A=\mathcal{S}$ by simplicity; similarly $\mathcal{C}A'=\mathcal{S}$. Thus there exist $B,B'\in\mathcal{C}$ with $A'=BA$ and $A=B'A'$. Since every row of $BA$ is a linear combination of the rows of $A$, we get $\sigma_r(A')\subseteq\sigma_r(A)$, and symmetrically $\sigma_r(A)\subseteq\sigma_r(A')$. Hence $\sigma_r(A)=\sigma_r(A')$, which is nonzero since $A\neq 0$.
\end{proof}

\begin{proposition}\label{p1}
    Let $\mathcal{C}$ be semisimple, and write
    $\mathcal{C}=\bigoplus_{i=1}^r\mathcal{A}_i$ with $\mathcal{A}_i$ simple
    left $\mathcal{C}$-modules. Then
    \[
    \mathcal{C} \text{ is row-intersecting} \iff r=1.
    \]
\end{proposition}

\begin{proof}
    If $r=1$, then $\mathcal{C}$ is a simple left $\mathcal{C}$-module, hence it is row-intersecting by Lemma~\ref{lemma:simple-rowspan} applied to $\mathcal{S}=\mathcal{C}$.

    Conversely, suppose $r\geq 2$. The decomposition
    $\mathcal{C}=\bigoplus_{i=1}^r\mathcal{A}_i$ yields a decomposition of the
    identity
    \[
    1_{\mathcal{C}}=e_1+\cdots+e_r,
    \]
    where $e_1,\ldots,e_r$ are nonzero pairwise orthogonal idempotents with
    $\mathcal{A}_i=\mathcal{C}e_i$ (see for example \cite{DrozdKirichenko}).
    By Lemma~\ref{lemma:orth-idemp}, $\sigma_r(e_1)\cap\sigma_r(e_2)=\{0\}$,
    so $\mathcal{C}$ is not row-intersecting.
\end{proof}

\begin{remark}\label{rmk:field}
    If $r=1$, then $\mathcal{C}$ has no nonzero proper left ideals, so it is a
    finite division algebra, hence a field by Wedderburn's little theorem.
    Thus a semisimple code is row-intersecting if and only if it is a field.
\end{remark}

\begin{example}
\begin{enumerate}
    \item Over $\mathbb{F}_2$, consider
    \[
    \mathcal{C}=\{0,\,A,\,B,\,A+B\},
    \qquad
    A=\begin{pmatrix} 1&1&0\\ 0&1&1\\ 1&0&1 \end{pmatrix},
    \quad
    B=\begin{pmatrix} 1&0&1\\ 1&1&0\\ 0&1&1 \end{pmatrix},
    \quad
    A+B=\begin{pmatrix} 0&1&1\\ 1&0&1\\ 1&1&0 \end{pmatrix}.
    \]
    One checks that $\mathcal{C}$ is closed under multiplication, with
    identity $1_{\mathcal{C}}=A+B$, and that $\mathcal{C}\cong\mathbb{F}_4$.
    Hence it is a simple module, and all its nonzero codewords have the same
    row space, namely the even-weight code
    $\{x\in\mathbb{F}_2^3 : x_1+x_2+x_3=0\}$.

        \item Over $\mathbb{F}_q$, consider the block-diagonal matrices
    \[
    \mathcal{C}=\left\{
    \begin{pmatrix} M & 0\\ 0 & e \end{pmatrix}
    \;:\; M\in M_2(\mathbb{F}_q),\ e\in\mathbb{F}_q
    \right\}\subseteq\mathbb{F}_q^{3\times 3}.
    \]
    As an algebra, $\mathcal{C}\cong M_2(\mathbb{F}_q)\times\mathbb{F}_q$, so it
    is semisimple. As a left $\mathcal{C}$-module, it decomposes into three
    simple modules: the two columns of $M_2(\mathbb{F}_q)$ and $\mathbb{F}_q$.
    Hence $\mathcal{C}$ is not row-intersecting by Proposition~\ref{p1}. Indeed, the idempotents
    $E_{1,1},E_{2,2}\in\mathcal{C}$ already have trivially intersecting row spaces.
\end{enumerate}
\end{example}

The aim of what follows is to reduce the problem to the study of simple or
semisimple algebras, which are much better understood. To do so, one can take
the quotient of $\mathcal{C}$ by a well-chosen ideal, or look for semisimple
submodules within $\mathcal{C}$.

Let us start with the first idea.

\begin{lemma}[Corollary 3.3.9 in \cite{DrozdKirichenko}]\label{lemma:lift}
    Let $\bar{\mathcal{C}}=\mathcal{C}/{\rm Rad}(\mathcal{C})$ and let
    $1_{\bar{\mathcal{C}}}=\bar e_1+\cdots+\bar e_r$ be a decomposition of the
    identity of $\bar{\mathcal{C}}$ into pairwise orthogonal idempotents. Then
    there exists a decomposition $1_{\mathcal{C}}=e_1+\cdots+e_r$ of the
    identity of $\mathcal{C}$ into pairwise orthogonal idempotents such that
    $\bar e_i=e_i+{\rm Rad}(\mathcal{C})$ for all $i\in\{1,\ldots,r\}$.
\end{lemma}

\begin{proposition}\label{prop:primary}
    If $\mathcal{C}$ is row-intersecting, then
    $\mathcal{C}/{\rm Rad}(\mathcal{C})$ is a field.
\end{proposition}

\begin{proof}
    Write $\bar{\mathcal{C}}=\mathcal{C}/{\rm Rad}(\mathcal{C})$, which is
    semisimple by Remark~\ref{r1}, and decompose it as
    $$\bar{\mathcal{C}}=\bigoplus_{i=1}^r\mathcal{A}_i$$ into simple left
    $\bar{\mathcal{C}}$-modules. As in Proposition~\ref{p1}, this gives a
    decomposition $$1_{\bar{\mathcal{C}}}=\bar e_1+\cdots+\bar e_r$$ into
    nonzero pairwise orthogonal idempotents. Suppose, for a contradiction,
    that $r\geq 2$. By Lemma~\ref{lemma:lift}, it lifts to a decomposition
    $1_{\mathcal{C}}=e_1+\cdots+e_r$ into pairwise orthogonal idempotents
    of $\mathcal{C}$. These are nonzero, since their images $\bar e_i$ are
    nonzero. By Lemma~\ref{lemma:orth-idemp},
    $\sigma_r(e_1)\cap\sigma_r(e_2)=\{0\}$, so $e_1,e_2$ are nonzero codewords
    with trivially intersecting row spaces, contradicting the assumption that
    $\mathcal{C}$ is row-intersecting. Hence $r=1$, and $\bar{\mathcal{C}}$ is
    a field (as in Remark~\ref{rmk:field}).
\end{proof}

\begin{example}\label{ex:radfield}
    Over $\mathbb{F}_q$, consider
    \[
    \mathcal{C}=\left\{
    \begin{pmatrix} a&0&0\\ b&a&0\\ c&0&a \end{pmatrix}
    \;:\; a,b,c\in\mathbb{F}_q
    \right\}.
    \]
    The code $\mathcal{C}$ is row-intersecting, because every nonzero $A\in\mathcal{C}$ satisfies $(1,0,0)\in\sigma_r(A)$:
    \begin{itemize}
        \item if $a\neq 0$, then $A$ is invertible and $\sigma_r(A)=\mathbb{F}_q^3$;
        \item if $a=0$, then $(b,c)\neq(0,0)$, and the nonzero rows of $A$ are all proportional to $(1,0,0)$.
    \end{itemize}
    Denote by $E_{i,j}$ the elementary matrices. The radical is the set of matrices with $a=0$:
    \[
    {\rm Rad}(\mathcal{C})=\langle E_{2,1},E_{3,1}\rangle_{\mathbb{F}_q}.
    \]
    Indeed, this is an ideal of $\mathcal{C}$ whose square is zero, and the quotient is a field. More precisely, the map $A\mapsto a$ induces an isomorphism
    \[
    \mathcal{C}/{\rm Rad}(\mathcal{C})\cong\mathbb{F}_q,
    \]
    which is a field, as stated in Proposition~\ref{prop:primary}.
\end{example}

\begin{remark}
    By Proposition~\ref{prop:primary}, the fact that $\mathcal{C}/{\rm Rad}(\mathcal{C})$ is a field is a necessary condition for an algebra $\mathcal{C}$ to be row-intersecting, so we may restrict our search to such algebras. However, the condition is not sufficient. To see this, consider the transpose of the code of Example~\ref{ex:radfield}:
    \[
    \mathcal{C}^{\top}=\left\{
    \begin{pmatrix} a&b&c\\ 0&a&0\\ 0&0&a \end{pmatrix}
    \;:\; a,b,c\in\mathbb{F}_q
    \right\}.
    \]
    Transposition is an anti-isomorphism of algebras, so it maps the radical of $\mathcal{C}$ onto the radical of $\mathcal{C}^{\top}$:
    \[
    {\rm Rad}(\mathcal{C}^{\top})=\langle E_{1,2},E_{1,3}\rangle_{\mathbb{F}_q},
    \]
    and $\mathcal{C}^{\top}/{\rm Rad}(\mathcal{C}^{\top})\cong\mathbb{F}_q$ is still a field. However, $\mathcal{C}^{\top}$ is not row-intersecting, since
    \[
    \sigma_r(E_{1,2})\cap\sigma_r(E_{1,3})
    =\langle(0,1,0)\rangle_{\mathbb{F}_q}\cap\langle(0,0,1)\rangle_{\mathbb{F}_q}=\{0\}.
    \]
\end{remark}

Let us now turn to the second idea: looking for a semisimple structure within $\mathcal{C}$.

\begin{theorem}\label{th:soc}
    The code $\mathcal{C}$ is row-intersecting if and only if ${\rm Soc}(\mathcal{C})$ is row-intersecting.
\end{theorem}

\begin{proof}
    Since ${\rm Soc}(\mathcal{C})\subseteq\mathcal{C}$, if $\mathcal{C}$ is row-intersecting then so is ${\rm Soc}(\mathcal{C})$.

    Conversely, assume that ${\rm Soc}(\mathcal{C})$ is row-intersecting. We first show that for every $B\in\mathcal{C}\setminus\{0\}$ there exists $A\in{\rm Soc}(\mathcal{C})\setminus\{0\}$ such that
    \[
    \sigma_r(A)\subseteq\sigma_r(B).
    \]
    Since $\mathcal{C}$ is unital, $B=1_{\mathcal{C}}B\in\mathcal{C}B$, so $\mathcal{C}B$ is a nonzero left ideal of $\mathcal{C}$. Being finite-dimensional, it contains a minimal left ideal $I$, and $I\subseteq{\rm Soc}(\mathcal{C})$ by definition of the socle. Take any $A\in I\setminus\{0\}$. Since $A\in\mathcal{C}B$, we can write $A=XB$ for some $X\in\mathcal{C}$. Every row of $XB$ is a linear combination of the rows of $B$, hence $\sigma_r(A)\subseteq\sigma_r(B)$.

    Now let $B,B'\in\mathcal{C}\setminus\{0\}$, and let $A,A'\in{\rm Soc}(\mathcal{C})\setminus\{0\}$ be as above. Since ${\rm Soc}(\mathcal{C})$ is row-intersecting,
    \[
    \{0\}\neq\sigma_r(A)\cap\sigma_r(A')\subseteq\sigma_r(B)\cap\sigma_r(B').
    \]
    Hence $\mathcal{C}$ is row-intersecting.
\end{proof}

\begin{remark}
    Beyond providing constructions, this theorem makes it more efficient to check whether such a code is row-intersecting. Indeed, the socle can be computed effectively (see for example in \cite{brooksbank2008testing}). After doing so, one can list all simple submodules of the socle. By Lemma~\ref{lemma:simple-rowspan}, each simple submodule has a constant rowspan among its nonzero matrices. One could then take one representative per submodule and compare their rowspans. The code is row-intersecting if and only if the rowspans of these representatives pairwise intersect.
    \end{remark}

\begin{corollary}\label{c1}
    If ${\rm Soc}(\mathcal{C})$ is simple, then $\mathcal{C}$ is row-intersecting.
\end{corollary}

\begin{proof}
    By Lemma~\ref{lemma:simple-rowspan}, ${\rm Soc}(\mathcal{C})$ is row-intersecting. The conclusion follows from Theorem~\ref{th:soc}.
\end{proof}

We list here three examples of row-intersecting codes with different socle structures.

\begin{example}
Let $\mathcal{C}=\mathbb{F}_q \left[\begin{pmatrix}
            0&1&0\\0&0&1\\0&0&0
        \end{pmatrix}\right]= \left\{\begin{pmatrix}
            a&b&c\\0&a&b\\0&0&a
        \end{pmatrix}; \: a,b,c\in \mathbb{F}_q\right\}$. We have \[{\rm Soc}(\mathcal{C})=\left< \begin{pmatrix}
            0&0&1\\0&0&0\\0&0&0
        \end{pmatrix}\right>_{\mathbb{F}_q},\] which is simple, so that the code $\mathcal{C}$ is row-intersecting.
\end{example}

\begin{example}\label{ex1}
Consider again the code $\mathcal{C}$ of Example~\ref{ex:radfield}. Since ${\rm Rad}(\mathcal{C})^2=0$, we have ${\rm Rad}(\mathcal{C})\subseteq{\rm Soc}(\mathcal{C})$, and since $\mathcal{C}$ is not semisimple and ${\rm Rad}(\mathcal{C})$ has codimension $1$, we get
\[
{\rm Soc}(\mathcal{C})={\rm Rad}(\mathcal{C})=\langle E_{2,1},E_{3,1}\rangle_{\mathbb{F}_q}.
\]
The socle has dimension $2$ and is not simple: since ${\rm Rad}(\mathcal{C})$ acts trivially on it, each of its $1$-dimensional subspaces is a submodule. However, all its nonzero elements have row space $\langle(1,0,0)\rangle_{\mathbb{F}_q}$, so it is row-intersecting, and so is $\mathcal{C}$ by Theorem~\ref{th:soc}.
\end{example}

\begin{example}\label{ex2}
Let $\mathcal{C}=\mathbb{F}_2\left[\begin{pmatrix}
            1&1&1&0&1\\1&1&0&1&1\\1&0&1&1&1\\0&1&1&1&1\\0&0&0&0&1
        \end{pmatrix}, \begin{pmatrix}
            0&1&1&1&1\\1&0&1&1&0\\1&1&0&1&0\\1&1&1&0&1\\0&0&0&0&1
        \end{pmatrix}\right]$. This is a commutative algebra of dimension $3$, and
        $${\rm Soc}(\mathcal{C})={\rm Rad}(\mathcal{C})=\left<\begin{pmatrix}
            0&1&1&0&1\\1&0&0&1&1\\1&0&0&1&1\\0&1&1&0&1\\0&0&0&0&0
        \end{pmatrix},\begin{pmatrix}
            1&1&1&1&1\\1&1&1&1&0\\1&1&1&1&0\\1&1&1&1&1\\0&0&0&0&0
        \end{pmatrix}\right>_{\mathbb{F}_2}.$$
        Since ${\rm Rad}(\mathcal{C})^2=0$, each $1$-dimensional subspace of the socle is a submodule, so the socle is not simple. One checks that the row spaces of its nonzero elements pairwise intersect, so ${\rm Soc}(\mathcal{C})$ is row-intersecting, and so is $\mathcal{C}$ by Theorem~\ref{th:soc}. Note that $\mathcal{C}$ has minimum distance $d=2$ (it can be easily checked), so it does not satisfy the condition $2d>n$ of Proposition~\ref{prop:min}. 
\end{example}

The main difference between the two previous examples is that in Example \ref{ex1} the intersection of the row supports of all nonzero codewords is non-trivial, whereas the code in Example \ref{ex2} does not have this property (that is, the global intersection is trivial). The first situation seems to be quite common: in our computer searches, we considered the algebras generated over $\mathbb{F}_2$ by subgroups of ${\rm GL}_n(\mathbb{F}_2)$, up to conjugacy, for $n\leq 6$, as well as algebras generated by random matrices, and the global intersection was frequently non-trivial among the row-intersecting ones.\\ 

The simplicity of the socle gives us a sufficient condition. There is a particular case in which it can also be necessary.

\begin{proposition}
    Let $\mathcal{A}$ be an algebra of dimension $n$ and $\mathcal{C}$ be its regular representation. Then $\mathcal{C}$, seen as a left module over itself, is row-intersecting if and only if ${\rm Soc}(\mathcal{C})$ is simple. 
\end{proposition}

\begin{proof}
    If ${\rm Soc}(\mathcal{C})$ is simple, then $\mathcal{C}$ is row-intersecting by Corollary~\ref{c1}. Conversely, write ${\rm Soc}(\mathcal{C})=\mathcal{C}A_1\oplus \ldots \oplus \mathcal{C}A_r$ as a direct sum of simple submodules. Let us assume that $r\geq 2$ and let us show that $\sigma_r(M_{A_1})\cap \sigma_r(M_{A_2})=\{0\}$.

    Let $B_1,\ldots,B_n$ be a $\mathbb{F}_q-$basis of $\mathcal{C}$. We define
        \[
\begin{aligned}
\kappa \colon \mathcal{C} &\xrightarrow{\sim} \mathbb{F}_q^{n} \\
A &\mapsto (\lambda_1,\ldots,\lambda_n)
\end{aligned}
\]
such that $A=\sum_{i=1}^n \lambda_iB_i$.

By construction, the $j$-th row of $M_{A_i}$ is given by $\kappa(B_jA_i)$, we then have:
$$\kappa^{-1}(\sigma_r(M_{A_i}))=\mathcal{C}A_i$$
for all $i\in \{1,\ldots,r\}$. Since the sum is direct, we have $\mathcal{C}A_1\cap \mathcal{C}A_2=\{0\}$, and since $\kappa$ is an isomorphism, we get: $$\sigma_r(M_{A_1})\cap \sigma_r(M_{A_2}) =\{0\}$$ 
Hence the code is not row-intersecting. 
\end{proof}

\begin{remark}
    Studying row-intersecting codes through the regular representation gives us a strictly algebraic criterion. However, it restricts the dimension of the algebras to be equal to the size of the matrices. 
\end{remark}

\begin{theorem}\label{th:dim}
    Let $\mathcal{C}\subseteq \mathbb{F}_q^{n\times n}$ be a unital algebra of dimension $k$. If ${\rm Soc}(\mathcal{C})$ is simple, then $k\leq n$.
\end{theorem}

\begin{proof}
    Let us assume that $k >n$. We then have: $$\forall v\in \mathbb{F}_q^n, \: \dim \mathcal{C}\cdot v< \dim \mathcal{C}$$ since $\mathcal{C}\cdot v\subseteq \mathbb{F}_q^n$ of dimension $n$.

    Hence the annihilator of $v$,
\[
{\rm Ann}_{\mathcal{C}}(v):=\{A\in\mathcal{C}\;:\;Av=0\},
\]
which is the kernel of the $\F_q$-linear map $\mathcal{C}\to\F_q^n$, $A\mapsto Av$, is nonzero. Moreover, it is a left ideal of $\mathcal{C}$, since $(BA)v=B(Av)=0$ for every $B\in\mathcal{C}$ and $A\in{\rm Ann}_{\mathcal{C}}(v)$. By simplicity and minimality of the socle, we have: $${\rm Soc}(\mathcal{C})\subseteq{\rm Ann}_\mathcal{C}(v)$$
    Since this is true for all $v\in \mathbb{F}_q^n$, we even have: $${\rm Soc}(\mathcal{C})\subseteq \cap _{v\in \mathbb{F}_q^n}{\rm Ann}_\mathcal{C}(v)$$
    However ${\rm Soc}(\mathcal{C}) \neq \{0\}$, which would mean that $ \cap _{v\in \mathbb{F}_q^n}{\rm Ann}_\mathcal{C}(v) \neq \{0\}$.

    This would imply that there exists a nonzero matrix whose kernel is equal to $\mathbb{F}_q^n$, hence the absurdity. We have then proven that $k\leq n$.
\end{proof}

\begin{remark}

Constructing row-intersecting codes that do not verify the sufficient condition $2d>n$ mentioned in Proposition \ref{prop:min} is nontrivial.

As mentioned in Remark~\ref{rmk:constructions}, avoiding this condition can be desirable.

The same reasoning applies to a second sufficient condition: that the socle of the module is simple. Again, studying row-intersecting codes through algebras with a simple socle would reveal properties of the simplicity of the socle rather than of the intersecting structure itself. The remaining case is therefore the one where the code is row-intersecting but the socle of the module is not simple. According to Theorem \ref{th:dim}, it is the only case where we can construct row-intersecting codes with dimension greater than the size of its matrices. 
\end{remark}

\vspace{1em}

Let us give an example of a construction avoiding both sufficient conditions. 

\vspace{1em}
\begin{example}\label{ex:block}
    
Let $\mathcal{G}\subseteq\mathbb{F}_q^{n\times n}$ be an $[n\times n,k,d]_q$-code.

Let $$\mathcal{C}=\left<\begin{pmatrix}
        I_n&A\\0&I_n
    \end{pmatrix}; \: A\in \mathcal{G}\right>_{\mathbb{F}_q}=\left\{\begin{pmatrix}
        \lambda I_n&A\\0&\lambda I_n
    \end{pmatrix}; \: \lambda\in\mathbb{F}_q,\ A\in \mathcal{G}\right\}\subseteq\mathbb{F}_q^{2n\times 2n}.$$

    The code $\mathcal{C}$ has dimension $k+1$ and minimum distance $d$: its nonzero non-invertible elements are the matrices with $\lambda=0$ and $A\neq 0$, whose rank is $\rk(A)$. It is stable through multiplication. Indeed, $\forall A,B \in \mathcal{G}$, $$\begin{pmatrix}
        I_n&A\\0&I_n
    \end{pmatrix} \begin{pmatrix}
        I_n &B\\0&I_n
    \end{pmatrix}=\begin{pmatrix}
        I_n&A+B\\0&I_n
    \end{pmatrix}\in \mathcal{C}.$$
    
    As a left-module, $\mathcal{C}$ has a unique maximal ideal, given by $\left\{\begin{pmatrix}
        0&A\\0&0
    \end{pmatrix}; \: A\in \mathcal{G}\right\}$. This ideal is the Jacobson radical of $\mathcal{C}$. Let us note that $$\forall A,B\in \mathcal{G}, \: \begin{pmatrix}
        0&A\\0&0
    \end{pmatrix} \begin{pmatrix}
        0&B\\0&0
    \end{pmatrix}=0.$$ 
    Hence ${\rm Rad}(\mathcal{C})^2=0$, so ${\rm Rad}(\mathcal{C})\subseteq {\rm Soc}(\mathcal{C})$ by the characterization of the socle recalled in Section~\ref{Sec:prelim}. Moreover, ${\rm Soc}(\mathcal{C})\neq\mathcal{C}$, since otherwise $\mathcal{C}$ would be semisimple and ${\rm Rad}(\mathcal{C})$ would be zero. As ${\rm Rad}(\mathcal{C})$ has codimension $1$ in $\mathcal{C}$, we get ${\rm Soc}(\mathcal{C})={\rm Rad}(\mathcal{C})$.
    Since $\sigma_r\begin{pmatrix} 0&A\\0&0 \end{pmatrix}=\{(0,x) : x\in\sigma_r(A)\}$, Theorem \ref{th:soc} gives $$\mathcal{C} \text{ is row-intersecting} \iff \mathcal{G} \text{ is row-intersecting.}$$

    For example, if $\mathcal{G}$ verifies $2d>n$, then it is row-intersecting. To maximize its dimension, one can consider an MRD code. If $n$ is even and $d=n/2+1$, then $\mathcal{G}$ is an $[n\times n, n^2/2, n/2+1]_q$-code. 

    We have constructed $\mathcal{C}$, a $[2n\times 2n, n^2/2+1, n/2+1]_q$-code. It does not verify either of the two sufficient conditions. 
\end{example}

\begin{remark}\label{rmk:bridge}
Let us compare the geometric and the algebraic points of view on the parameters. For a row-intersecting algebra $\mathcal{C}\subseteq\mathbb{F}_q^{n\times n}$ of dimension $k$ and minimum distance $d$, Theorem~\ref{thm:md} gives $k\leq nd$. When ${\rm Soc}(\mathcal{C})$ is simple, Theorem~\ref{th:dim} sharpens this to $k\leq n$, independently of $d$. On the other hand, the code of Example~\ref{ex:block} shows that row-intersecting algebras with a non-simple socle can have dimension much larger than the size of their matrices: it has parameters $[2n\times 2n,\,n^2/2+1,\,n/2+1]_q$, so that $k$ grows quadratically in $n$, while still satisfying the geometric bound $k\leq 2n\cdot d$.
\end{remark}

\begin{remark}\label{rmk:modules}
Several results of this section do not require $\mathcal{C}$ to be closed under multiplication. Let $\mathcal{C}\subseteq\F_q^{m\times n}$ be any code and let $R$ be a unital subalgebra of its left idealizer $$\{X\in\F_q^{m\times m}:X\mathcal{C}\subseteq\mathcal{C}\};$$ for instance, $R=\F_q[G]$ for a group $G\leq{\rm GL}_m(\F_q)$ of left automorphisms of $\mathcal{C}$. Then $\mathcal{C}$ is a left $R$-module, and in what follows ${\rm Soc}(\mathcal{C})$ denotes its socle as such. Since $\sigma_r(XA)\subseteq\sigma_r(A)$ for every $X\in\F_q^{m\times m}$, the proofs of Lemma~\ref{lemma:simple-rowspan}, Theorem~\ref{th:soc} and Corollary~\ref{c1} carry over, and the proof of Theorem~\ref{th:dim} gives $k\leq m$ whenever ${\rm Soc}(\mathcal{C})$ is simple. On the contrary, Propositions~\ref{p1} and~\ref{prop:primary} rely on the fact that idempotents of $\mathcal{C}$ are codewords. Note that if $R$ is semisimple, e.g.\ $R=\F_q[G]$ with $\gcd(|G|,q)=1$ (see Maschke's Theorem \cite[Theorem~7.1.1]{DrozdKirichenko}), then ${\rm Soc}(\mathcal{C})=\mathcal{C}$ and Theorem~\ref{th:soc} gives no information; the interesting case is therefore the modular one. A systematic study of this more general setting is left for future work.
\end{remark}

\vspace{1.5em}

Let us end this section by considering a particular case for which the algebraic structure is completely understood.

\subsection{The case $\mathcal{C}=\mathbb{F}_q[B], \: B\in \mathbb{F}_q^{n\times n}$.}

\begin{proposition}\label{prop:params}
Let $B\in\mathbb{F}_q^{n\times n}$ and $\mu_B(x)=\prod_{i=1}^{r}P_i^{e_i}(x)\in \mathbb{F}_q[x]$ be its minimal polynomial, where the $P_i(x)$ are
distinct irreducible polynomials.

Let $l_i:=\deg(P_i(x))$, and let $P_i^{\alpha_{i,1}}(x),\dots,P_i^{\alpha_{i,s_i}}(x)$,
with $\alpha_{i,1}\leq\dots\leq\alpha_{i,s_i}=e_i$, be the invariant factors of
the restriction $B_i$ of $B$ to $\ker(P_i^{e_i}(B))$. Define
\[
d_i:=l_i\times\#\bigl\{j\in\{1,\dots,s_i\}\;:\;
\alpha_{i,j}=e_i\bigr\}.
\]
Then the code $\mathbb{F}_q[B]\subseteq\mathbb{F}_q^{n\times n}$ is an $\bigl[n\times n, \deg(\mu_B), \min_{1\leq i\leq r} (d_i)\bigr]_q$-code.
\end{proposition}

\begin{proof}
We have $\mathbb{F}_q[B]\cong \mathbb{F}_q[x]/(\mu_B(x))$. Since $\{1, x,\ldots,x^{\deg(\mu_B)-1}\}$ is a $\mathbb{F}_q$-basis of $\mathbb{F}_q[x]/(\mu_B(x))$, the dimension of the algebra $\mathbb{F}_q[B]$ as an $\mathbb{F}_q$-vector space is equal to $\deg(\mu_B(x))$.

Let us now express the minimum distance. By the primary decomposition theorem, one can write
\[
\mathbb{F}_q^n=\bigoplus_{i=1}^{r}\ker\bigl(P_i^{e_i}(B)\bigr),
\]
Each $\ker\bigl(P_i^{e_i}(B)\bigr)$ is $B$-stable so, taking a basis adapted to this decomposition, there exists $Q\in GL_n(\mathbb{F}_q)$ such that $Q^{-1}BQ=\operatorname{diag}(B_1,\dots,B_r)$.
Since $M\mapsto Q^{-1}MQ$ preserves the rank, we may assume that
$B=\operatorname{diag}(B_1,\dots,B_r)$. Then for every $f\in\mathbb{F}_q[x]$,
$$f(B)=\operatorname{diag}\bigl(f(B_1),\dots,f(B_r)\bigr),\qquad \operatorname{rank} f(B)=\sum_{i=1}^r \operatorname{rank} f(B_i).$$

Moreover, for all $1\leq i\leq r$, the minimal polynomial of $B_i$ is $\mu_{B_i}(x)=P_i^{e_i}(x)$. In particular, choosing $f(x)=\prod_{j\neq i}P_j^{e_j}(x)$ vanishes on all $B_j$ but $f(B_i)\neq 0$. This shows that the minimum distance $d$ of $\mathbb{F}_q[B]$ is equal to: $$d=\min_{1\leq i\leq r}(\underbrace{\min_{A\in \mathbb{F}_q[B_i]\backslash\{0\}}(\rk (A))}_{d(\mathbb{F}_q[B_i])})$$

We can now express the minimum distance $d(\mathbb{F}_q[B_i])$ of each $\mathbb{F}_q[B_i]$. 
Let $i\in\{1,\dots,r\}$ and consider the invariant factors of $B_i$. Since
$P_i(x)$ is irreducible, they are
\[
\{P_i^{\alpha_{i,1}}(x),\dots,P_i^{\alpha_{i,s_i}}(x)\},
\qquad \alpha_{i,s_i}=e_i .
\]

Let $C_{P_i^{\alpha_{i,j}}}$ denote the companion matrix of $P_i^{\alpha_{i,j}}(x)$.
By the Frobenius (rational canonical form) theorem, there exists $R\in {\rm GL}_{n_i}(\mathbb{F}_q)$ such that $$R^{-1}B_iR={\rm diag}(C_{P_i^{\alpha_{i, 1}}},\ldots,C_{P_i^{\alpha_{i, s_i}}}).$$ 

Each companion matrix can itself be decomposed. Let $1\leq j\leq s_i$ and $C_{P_i}$ denote the companion matrix of $P_i$. There exists $S\in{\rm GL}_{l_i\alpha_{i,j}}(\mathbb{F}_q)$ such that $$S^{-1}C_{P_i^{\alpha_{i,j}}}S=\underbrace{
\begin{pmatrix}
    C_{P_i} & I_{l_i} & & (0)\\
    & \ddots & \ddots & \\
    (0) & & C_{P_i} & I_{l_i}
\end{pmatrix}
}_{\alpha_{i,j}\text{ blocks}}\in \mathbb{F}_q^{l_i\cdot\alpha_{i,j}\times l_i\cdot \alpha_{i,j}}
$$

We have: $$\min_{A\in \mathbb{F}_q[C_{P_i^{\alpha_{i,j}}}]\backslash\{0\}}(\rk(A))=l_i$$

Note that for all $g(x)\in \mathbb{F}_q[x]$, if $g(B_i)\neq 0$ then for all $j$ such that $\alpha_{i,j}=e_i$, $g(S^{-1}C_{P_i^{\alpha_{i,j}}}S)\neq0$. 

This show that $$d(\mathbb{F}_q[B_i]) \geq l_i\times \#\bigl\{j\in\{1,\dots,s_i\}\;:\;
\alpha_{i,j}=e_i\bigr\}$$

Furthermore, $\rk (P_i^{e_i-1}(B_i))=l_i\times \#\bigl\{j\in\{1,\dots,s_i\}\;:\;
\alpha_{i,j}=e_i\bigr\}$

Hence $$d(\mathbb{F}_q[B_i])=d_i=l_i\times \#\bigl\{j\in\{1,\dots,s_i\}\;:\;
\alpha_{i,j}=e_i\bigr\}.$$

To conclude, the minimum distance of $\mathbb{F}_q[B]$ is $\min_{1\leq i\leq r} (d_i)$.
\end{proof}

\vspace{1.5em}

Since codes are often considered up to equivalence, the following proposition shows that two codes of the form $\mathbb{F}_q[B]$ are equivalent if and only if they are conjugate as subalgebras of $\mathbb{F}_q^{n\times n}$.

\begin{proposition}
    Let $B,\Lambda\in \mathbb{F}_q^{n\times n}$. The codes $\mathbb{F}_q[B]$ and $\mathbb{F}_q[\Lambda]$ are equivalent  if and only if there exists $P\in {\rm GL}_n(\mathbb{F}_q)$ such that
    \[
    \mathbb{F}_q[\Lambda]=P\mathbb{F}_q[B]P^{-1}.
    \]
\end{proposition}

\begin{proof}
    If $\mathbb{F}_q[B]$ and $\mathbb{F}_q[\Lambda]$ are equivalent codes, there exist $P,Q\in {\rm GL}_n(\mathbb{F}_q)$ such that
    \[
    \mathbb{F}_q[\Lambda]=P\mathbb{F}_q[B]Q.
    \]
    
    In this case, since $P\mathbb{F}_q[B]Q=\mathbb{F}_q[\Lambda ]$ is a unital algebra, $I_n\in P\mathbb{F}_q[B]Q.$
    Thus, there exists $f(x)\in \mathbb{F}_q[x]$ such that $Pf(B)Q=I_n$, i.e. $f(B)=P^{-1}Q^{-1}$.
    
    Moreover,
    \[
    P^{-1}Q^{-1}=(QP)^{-1}\in\mathbb{F}_q[B]^\times 
    \Longrightarrow
    QP\in\mathbb{F}_q[B]
    \]
    Hence, there exists $h\in \mathbb{F}_q[x]$, $h(B)=QP$.

    Then $Q=h(B)P^{-1}$, and since $h(B)$ is a unit of
$\mathbb{F}_q[B]$ we have $\mathbb{F}_q[B]\,h(B)=\mathbb{F}_q[B]$. Therefore
\[
\mathbb{F}_q[\Lambda]=P\,\mathbb{F}_q[B]\,h(B)P^{-1}=P\,\mathbb{F}_q[B]\,P^{-1}.
\]

\end{proof}

\vspace{1.5em}

We can now conclude by characterizing which of these codes are row-intersecting.

\begin{proposition}\label{prop:Fqb}
    Let $B\in\mathbb{F}_q^{n\times n}$ and let $\mu_B(x)=\prod_{i=1}^{r}P_i^{e_i}(x)\in \mathbb{F}_q[x]$ be its minimal polynomial, where the $P_i(x)$ are distinct irreducible polynomials. 
    
    Then the code $\mathcal{C}=\mathbb{F}_q[B]$ is row-intersecting if and only if $r=1$. 
\end{proposition}

\begin{proof}
    Let $\mathcal{A}=\mathbb{F}_q[B]$. $$\mathcal{A}\cong \mathbb{F}_q[x]/(\mu_B(x))\cong \prod_{i=1}^r\mathbb{F}_q[x]/(P_i^{e_i}(x)).$$ Each factor $\mathcal{A}_i=\mathbb{F}_q[x]/(P_i^{e_i}(x))$ is a local algebra. $\mathcal{A}$ acts on $\mathbb{F}_q[x]/(P_i^{e_i}(x))$ through the quotient map $\mathcal{A}\twoheadrightarrow\mathbb{F}_q[x]/(P_i^{e_i}(x))$. This allows us to view $\mathbb{F}_q[x]/(P_i^{e_i}(x))$ as a left $\mathcal{A}$-module whose submodules form the chain $(\overline{P_i(x)})\supset\dots\supset(\overline{P_i^{e_i}(x)})=\overline{0}$. Its unique simple submodule is $(\overline{P_i^{e_i-1}(x)})$. Since $\mathcal{A}\cong\prod_i \mathcal{A}_i$, the simple submodules of $\mathcal{A}$ are the $(\overline{P_i^{e_i-1}(x)})$. Hence, as $\mathcal{A}$-modules,
$$\operatorname{Soc}(\mathcal{A})=\bigoplus_{i=1}^r (\overline{P_i^{e_i-1}(x)}),
  \qquad
  (\overline{P_i^{e_i-1}(x)})\cong \mathbb{F}_q[x]/(P_i(x))\cong\mathbb{F}_{q^{\deg P_i}}.$$

By Corollary \ref{c1}, if $\mu_B(x)$ has exactly one factor, the socle is simple and the code is row-intersecting.

Suppose that $r\geq 2$.

There exists $R\in {\rm GL}_n(\mathbb{F}_q)$ such that $R^{-1}BR={\rm diag}(B_1,\ldots,B_r)$ and $\mu_{B_i}(x)=P_i^{e_i}(x)$. 

    Setting $F_j=\prod _{i\neq j}P_i^{e_i}(x)$, we obtain $$F_j(B)={\rm diag}(0,\ldots,0,(*),0,\ldots,0)$$
    where the $(*)$ block is in the j-th position and is nonzero by minimality of $\mu_B(x)$.

    Consequently, $$\sigma_r(F_j(B))\cap \sigma_r(F_{j'}(B))=\{0\} \text{  if } j\neq j'.$$

    and the code is not row-intersecting.
\end{proof}

\begin{remark}
    In this case, $\mathbb{F}_q[B]$ is row intersecting if and only if its socle is simple.
\end{remark}

\begin{example}
    \begin{enumerate}
        \item Let $B=\begin{pmatrix}0&1&1&0\\1&0&1&1\\0&0&1&0\\1&0&0&0\end{pmatrix}\in\mathbb{F}_2^{4\times 4}$. Then $$\mu_B=x^4+x^3+x^2+1=(x^3+x+1)(x+1),$$ with two distinct irreducible factors. By Proposition~\ref{prop:params}, $\mathbb{F}_2[B]$ is a $[4\times 4,4,1]_2$ code, and it is not row-intersecting by Proposition~\ref{prop:Fqb}.
        \item Let $B=\begin{pmatrix}0&1&0&1\\0&0&1&0\\1&0&1&0\\0&0&1&1\end{pmatrix}\in\mathbb{F}_2^{4\times 4}$. Then $$\mu_B=x^4+x^2+1=(x^2+x+1)^2,$$ so by Proposition~\ref{prop:params} the code $\mathbb{F}_2[B]$ is a $[4\times 4,4,2]_2$ code, and it is row-intersecting by Proposition~\ref{prop:Fqb}. Note that $2d=n$, so the condition of Proposition~\ref{prop:min} is not satisfied. Its socle is
        \[
        {\rm Soc}(\mathbb{F}_2[B])=(B^2+B+I)\,\mathbb{F}_2[B]=\left\langle
        \begin{pmatrix}1&1&0&0\\1&1&0&0\\0&1&1&1\\1&0&1&1\end{pmatrix},
        \begin{pmatrix}0&1&1&1\\0&1&1&1\\1&0&1&1\\1&1&0&0\end{pmatrix}
        \right\rangle_{\mathbb{F}_2}\cong\mathbb{F}_4,
        \]
        and its three nonzero elements have rank $2$ and the same row space $\langle(1,1,0,0),(0,1,1,1)\rangle_{\mathbb{F}_2}$, which is contained in the row space of every nonzero codeword.
    \end{enumerate}
\end{example}

\section*{Statement on the use of AI tools}
The authors used Claude  Opus 5.5 (Anthropic) for proofreading, for drafting expository passages on standard material, and for checking the mathematical consistency of the manuscript; in this process it suggested some corrections and the formulation of a few auxiliary statements and short proofs. All suggestions were verified by the authors, who take full responsibility for the content.

\bibliographystyle{alpha}
\bibliography{articles}

\end{document}